\newcommand{\tc}{$T_{\text c}$}
\newcommand{\cci}{CeCoIn$_5$}
\newcommand{\cirin}{CeIrIn$_5$}
\newcommand{\crhin}{CeRhIn$_5$}
\newcommand{\sro}{Sr$_2$RuO$_4$}

\newcommand{\jphysc}{J.~Phys.:~Con.~Mat.~}

\documentclass[prl,twocolumn,showpacs,preprintnumbers,amsmath,amssymb]{revtex4}
\usepackage{graphicx}% Include figure files

\begin{document}
%\raggedbottom
%\preprint{APS/123-QED}

\title{\protect\cci\ -- a quantum critical superfluid? }

\author{S.~\"Ozcan}
\author{D.~M.~Broun}
\altaffiliation{Now at Dept. of Physics and Astronomy, University of
British Columbia, Vancouver B.C., Canada V6T 1Z1.}
\author{B.~Morgan}%
\author{R.~K.~W.~Haselwimmer}
\author{J.~R.~Waldram}
\address{Cavendish Laboratory, University of Cambridge, Madingley Road,
Cambridge, CB3 0HE, United Kingdom}

\author{J.~L.~Sarrao}
\address{Los Alamos National Laboratory, Los Alamos, New Mexico 87545, USA}%

\author{Saeid Kamal}
\author{C.~P.~Bidinosti}
\author{ P.~J.~Turner}
\address{Department of Physics and Astronomy, University of British
Columbia, Vancouver B.C., Canada V6T 1Z1}

\date{\today}

\begin{abstract}
We have made the first complete measurements of the London penetration depth $\lambda(T)$ of \cci, a quantum-critical
metal where superconductivity arises from a non-Fermi-liquid normal state.  Using a novel tunnel diode oscillator
designed to avoid spurious contributions to $\lambda(T)$, we have established the existence of intrinsic and anomalous
power-law behaviour at low temperature. A systematic analysis raises the possibility that the unusual observations are
due to an extension of quantum criticality into the superconducting state.
\end{abstract}
\pacs{74.25.NF, 74.70.Tx}% PACS, the Physics and Astronomy
\maketitle

A number of experiments on the recently discovered family of heavy fermion superconductors \crhin, \cirin\ and \cci\
point to the existence of a non-Fermi-liquid (NFL) metallic state in these compounds. Central to this conclusion is the
observation, in
\cirin\ and \cci, that two of the key parameters of a Fermi liquid --- the electronic heat capacity coefficient $\gamma =
C/T$ and Pauli  susceptibility $\chi$ --- increase on cooling and show no sign of entering a temperature-independent
Fermi-liquid regime \cite{kim}.   Similar behaviour is observed in other heavy-fermion materials \cite{varma} and is
understood in terms of magnetic fluctuations near a  zero-temperature critical point, where theory predicts either
$\gamma \sim -\ln T$ or $\gamma = \gamma_0 - AT^{1/2}$,  depending  on the dimensionality and nature of the
magnetism \cite{millis,moriya}.  What happens to such a system when it becomes  superconducting is at present an 
open question.  Do the Fermi liquid parameters continue to evolve within the superconducting state, or  does
superconductivity abort the approach to quantum criticality?  Also, would the absence of quantum-critical superfluidity
in  materials such as the cuprate superconductors rule out a zero-temperature critical point as the source of  their NFL
normal-state  behaviour? These are important issues, which we address in this Letter with the first complete and well
controlled measurements of the London penetration depth
$\lambda(T)$ of a superconductor known to be situated  near a magnetic quantum critical point.  Our novel oscillator
design and carefully characterised samples avoid extrinsic contributions to the $\lambda(T)$ signal, and a systematic
analysis leads us to consider new, NFL physics as the reason for the anomalous power-law behaviour observed in
$\lambda(T)$ in the low temperature limit.

The normal phase of \cci\ contains all the hallmarks of quantum criticality.  De Haas-van Alphen (dHvA) measurements reveal large
cyclotron masses that are strongly field dependent, exceeding 100~$m_0$ at low fields \cite{settai}.  The resistivity is reminiscent
of the cuprates, nearly linear in temperature below 20~K \cite{petrovic}. In magnetic fields large enough to suppress
superconductivity, $C/T \sim -\ln T$ down to the lowest temperatures \cite{petrovic,kim}.  From 2.5~K up to 100~K the
spin--lattice relaxation rate has the $T^{1/4}$ temperature dependence expected near an antiferromagnetic instability
\cite{kohori}, indicating proximity to a zero-temperature magnetic critical point.  This is supported by the fact that \cci\ is
a higher-density analogue of  \crhin, a material in which weak, ambient-pressure antiferromagnetism gives way to
non-$s$-wave superconductivity with \tc\ = 2.1~K at a pressure of 1.6~GPa \cite{hegger,sidorov}.

The superconducting state of \cci\ is also highly unconventional and appears to bear strong similarities to that of the
cuprate superconductors \cite{tsuei}. Measurements of specific heat and thermal conductivity $\kappa(T)$ reveal low
temperature power-law behaviour, similar to expectations for a pairing state with line nodes
\cite{petrovic,movshovich}.  A strong-power law temperature dependence is also observed in the low-temperature
microwave surface impedance \cite{ormeno}. The four-fold modulation of $\kappa$ by an angle-dependent basal-plane
magnetic field indicates that the line nodes lie along the [110] directions \cite{izawa}. This is confirmed by a
field-rotation study of dHvA oscillations, which {\em increase} in amplitude below $H_{c2}$ for fields in the [110]
direction \cite{settai}. In addition, observations of Pauli-limited superconductivity in $\kappa(H)$
\cite{izawa} and a Knight shift that decreases below \tc\ in all directions \cite{kohori} imply spin-singlet
superconductivity. Together the experiments strongly suggest that pairing occurs in a  $d_{x^2 - y^2}$ state.

\cci\ forms in a  tetragonal crystal structure with alternating layers of CeIn$_3$ and CoIn$_2$, and superconducts at
ambient pressure below $T_{\text c}= 2.25$~K.   The high-quality \cci\ crystals used in our experiment were grown by a
self-flux method in excess In \cite{petrovic,kim}. Although the starting materials are very pure to begin with (Ce:
99.99\%, Co: 99.9975\% and In: 99.9995\%), growth in excess In is expected to further refine the crystals.  These
naturally form as large $ab$-plane platelets, with mirror-like surfaces, and are ideally suited to high-frequency
measurements.  The high homogeneity and low defect level of the crystals are confirmed by our microwave
measurements, which show a sharp superconducting transition ($\Delta$\tc~$ < 30$~mK) and a low quasiparticle
scattering rate ($1/\tau$, the width of the conductivity spectrum $\sigma(\omega)$,
$\approx 2\times 10^{10}$~s$^{-1}$ at 1.2~K) \cite{turner}. We focus on data from one sample, a crystal with dimensions
$a\times b\times c = 1.38\times 1.37
\times 0.073$~mm$^3$.

The $\lambda(T)$ measurements reported here were made with a 130~MHz tunnel diode oscillator (TDO) operated in
$^4$He and dilution refrigerator cryostats over the temperature range 0.1~K to 9~K.  In addition, a thorough study of the
microwave conductivity $\sigma(\omega,T)$, obtained from surface  impedance measurements  from 1 to 75~GHz, was made on the 
same crystals used in the TDO experiments.  Although the microwave measurements form an important part of this work (as a sample
characterisation tool, for cross-checking and calibrating the TDO measurements, and as a means of determining the {\em absolute}
penetration depth) a detailed discussion of those results will be presented elsewhere \cite{broun}.  We focus here instead on
describing the distinctive features of our tunnel diode oscillator and the results obtained using it.

\begin{figure}
\includegraphics[scale=0.75]{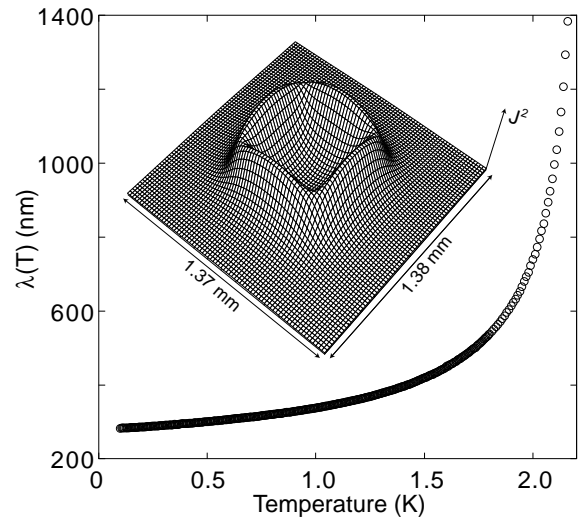}
\caption{\label{figure1} Temperature dependence of the penetration depth in
\cci.  Inset: A calculation
of the induced surface current density $J(x,y)$ in the \cci\ sample.  Note
that current  flows in a ring, well within the
boundaries of the crystal face.}
\end{figure}

A good description of  the general principles of operation of the TDO is given in Ref.~\onlinecite{Degrift1}, and
oscillators of this sort have been used by other workers to study unconventional superconductors
\cite{carrington,bonalde}.  The chief innovation in our apparatus is the novel geometry of the oscillator's probe
circuit --- a very small, self-resonant superconducting coil that acts as a local probe of the magnetic penetration depth. 
The miniature resonator has a square cross-section of side 0.5~mm and is wound from superconducting Nb wire on a
high purity sapphire former.  The high quality factor of the coil ($Q = 2.5\times10^5$ at 1.3\,K) allows the resonator to be
coupled inductively to the tunnel diode, with the mutual inductance backed-off until the diode only marginally  sustains
oscillation.  This results in a frequency stability  better than 1 part in
$10^9$ per hour and a field at the sample of $\sim 10$~nT.  The combination of a high-$Q$ resonator and low-power
tunnel diode keeps the total heat load of the oscillator below 2\,$\mu$W and allows the whole circuit to be cooled into
the mK temperature range.  In the experiments, a basal-plane face of the single-crystal sample is attached to one end of
the sapphire former with vacuum grease, about 50\,$\mu$m from the end of  the coil. In this geometry, the coil locally
induces currents that flow in a 0.5~mm wide ring in the centre of the crystal face, as shown in the inset of
Fig.~\ref{figure1}. This is important when studying the superfluid response of electrically anisotropic
materials such as \cci, for two reasons:  first, only basal-plane currents flow, eliminating contamination of the signal by
$c$-axis currents; secondly, this experimental geometry is inherently insensitive to nonlocal effects, as electrons in a
quasi-2D metal are {\em intrinsically} confined to move parallel to the basal plane
\cite{kosztin1}.

In our setup there is thermal contact between the sample and the resonator, with the result that the temperature of the
entire oscillator is swept during the experiments. Being able to operate in this mode has the potential advantage that both
resonator and sample can be embedded in a hydrostatic pressure medium, in principle allowing the investigation of other
quantum-critical systems, such as CePd$_2$Si$_2$ and CeIn$_3$ \cite{grosche}, which only superconduct under high
pressures. However, changing the temperature of the oscillator also introduces systematic errors into the frequency shift
signal. Fortunately, these are small over most of the temperature range and are highly reproducible, allowing them to be
accurately accounted for using background measurements made in the absence of the sample.
$\Delta\lambda(T)$ is obtained from the oscillator frequency-shift signal
$\Delta f_0 (T) = f_0(T) - f_0(T_{\text {base}})$ using the cavity perturbation approximation, where $\Delta \lambda
(T)= -\Gamma\Delta f_0 (T)$. Here $\Gamma = 6$~\AA/Hz is a temperature-independent geometric factor determined
empirically by comparison with penetration depth measurements made down to 1.2~K using a 1~GHz loop--gap
resonator \cite{hardy93}.

\begin{figure}
\includegraphics[scale=0.75]{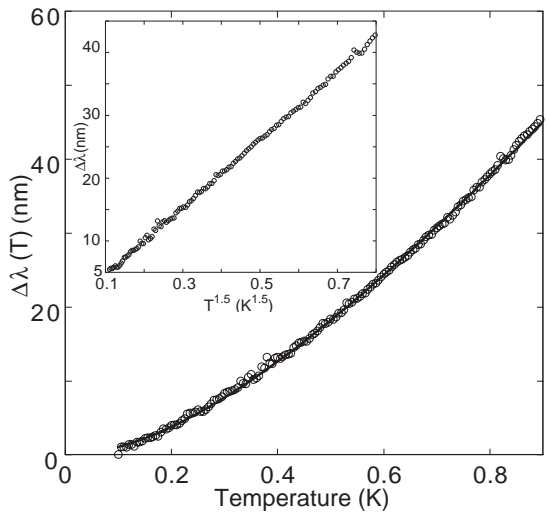}
\caption{\label{figure2} Two views of $\lambda(T)$ in the low-$T$ limit,
revealing a strong
$T$ dependence and the presence of gapless excitations.  The solid line is
a fit to the function
$\Delta\lambda(T) = A T^2/(T + T^\ast)$.  Inset: $\lambda(T)$ is also
well-described by a $T^{1.5}$ power law.}
\end{figure}

The absolute value of $\lambda(T)$ has been determined from measurements of the surface impedance $Z_{\text s} =
R_{\text s} + {\text i} X_{\text s}$, made at 5.5~GHz with a TE$_{011}$ mode dielectric resonator.  It is usually very
difficult to determine
$\lambda$ in this way, because only {\em shifts} in $X_{\text s}$ with temperature are experimentally accessible.
However, by carrying measurements of $\Delta X_{\text s}(T)$ up to temperatures where the electronic scattering rate is
much larger than the microwave frequency (i.e. for $T \ge 20$~K in \cci) we access the Hagen--Rubens limit where
$R_{\text s} \approx X_{\text s}$. This enables the unknown offset in $X_{\text s}$ to be determined and, for
sufficiently accurate $\Delta X_{\text s}(T)$ data, gives
$X_{\text s}(T)$ absolutely down to 1.2~K.  (Confidence in our procedure is enhanced by the fact that
$R_{\text s}$ and $X_{\text s}$ match between 20~K and 90~K, and that an analysis of the {\em departure} of $R_{\text
s}$ from $X_{\text s}$ below 20~K reveals a temperature-dependence of the optical effective mass that follows the $\ln
T$ behaviour of $C/T$ \cite{broun}.)  Having the $X_{\text s}(T)$ data is not enough: only in the low frequency limit, $\omega\tau
\ll 1$, does
$X_{\text s} = \omega\mu_0\lambda$; at higher frequencies thermally excited quasiparticles also contribute to
$X_{\text s}$
\cite{hosseini}.  We use bolometric measurements of $\sigma(\omega,T)$
\cite{turner} in the superconducting state to properly account for the quasiparticle contribution,  obtaining
$\lambda(1.2$~K$) = 3610$~\AA\ and $\lambda_0 = 2810$~\AA\ in the $T \rightarrow 0$ limit. This is a large
penetration depth, characteristic of a metal with heavily renormalised electrons.

\begin{figure}
\includegraphics[scale=0.75]{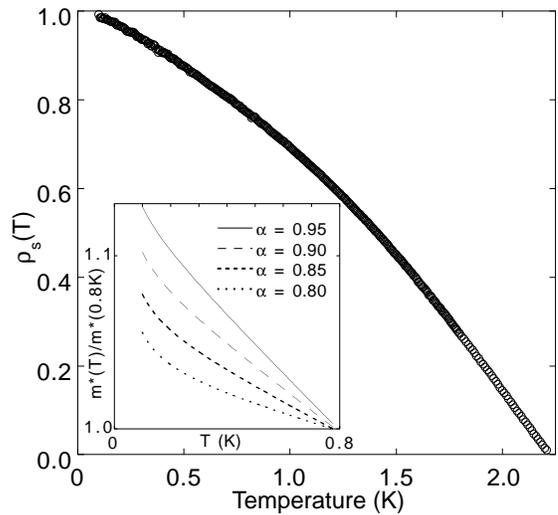}
\caption{\label{figure3} $\rho_{\text s}(T) = \lambda^2(0)/\lambda^2(T)$
for \cci, showing mean- field
behaviour near $T_{\text c}$ and a strong, but {\em nowhere linear}, $T$
dependence at low $T$.  Inset: $T$ dependence of
the average effective mass $m^\ast$ required to bring the observed
$\rho_{\text s}(T) \propto n_{\text s}(T)/m^\ast(T)$ into
accord with the expected form for line nodes, $n_{\text s}(T) = n_0(1 -
\alpha T/T_{\text c})$, for several values of $\alpha$. }
\end{figure}

Figure~\ref{figure1} shows the  absolute $\lambda(T)$ data measured with the tunnel diode oscillator down to 0.1~K.  In
Fig.~\ref{figure2}, a close-up view of the low-$T$ data reveals a strong temperature dependence, indicating low-lying
excitations and supporting the case for line nodes in the pairing state.   However,
$\lambda(T)$ does not have the simple linear $T$ dependence expected for a $d$-wave superconductor and observed in
the cuprates \cite{hardy93}: instead the inset to Fig.~\ref{figure2} reveals that $\lambda(T)$ is  better approximated by a
$T^{1.5}$ power law, down to 0.1~K.  (Similar curvature in $\lambda(T)$ has been inferred from 10~GHz $X_{\text
s}(T)$ measurements and attributed to disorder \cite{ormeno} .  In those measurements, made down to only 0.25~K,
$\omega\tau \approx 5$ at low $T$ and the clear connection between $X_{\text s}$ and the London penetration depth is
lost.)  Also relevant is the observation that $\kappa(T) \sim T^{3.37}$ below 0.2~K \cite{movshovich}.  The expected 
power law in the case of dilute strong-scattering impurities is $\kappa(T) \sim T^3$ \cite{graf}, and an interesting 
possibility is that the fractional power laws in
$\kappa(T)$ and $\lambda(T)$ might have the same origin.  In Fig.~\ref{figure3} the data are plotted as $\rho_{\text
s}(T) = \lambda^2(0)/\lambda^2(T)$.  As $\rho_{\text s}$ is proportional to the ratio of the superfluid density $n_{\text s}$ to
the effective mass $m^\ast$, it is expected to have a linear $T$ dependence in a $d$-wave superconductor, a geometric
consequence of the presence of line nodes in the energy gap.  This is not seen in our data, which show curvature over the
full temperature range.

We now consider possible explanations for the observed $T$-dependence of
$\lambda$, beginning with disorder.  Strong-scattering impurities in a
$d$-wave superconductor are known to induce a crossover in
$\lambda(T)$ from clean-limit, $T$-linear behaviour above $T^\ast\approx 0.56 \sqrt{n_{\text i} T_{\text F} T_{\text
c}}$ to a quadratic
$T$ dependence at low temperatures \cite{hirschfeld}. (Here $n_{\text i}$ is the density of impurities and $T_{\text F}$ the Fermi
temperature.) We have assessed this possibility by fitting the interpolation formula of Ref.~\onlinecite{hirschfeld},
$\Delta\lambda(T) \propto T^2/(T + T^\ast)$, to  our data in Fig.~\ref{figure2}, and obtain $T^\ast = 0.3$~K.  Using
$T_{\text F} = 50$~K (from specific heat
\cite{kim}) we infer from $T^\ast$ that $n_{\text i} = 0.26$\%.  This is over an order of magnitude greater than the
density of impurities in  our starting materials.  Thermal conductivity experiments suggest the level of strong-scattering
defects in our crystals is actually {\em much} lower: in $\kappa(T)$, $T^\ast < 30$~mK, implying $n_{\text i} <
26$~ppm \cite{movshovich}.

Alternatively, Kosztin and Leggett \cite{kosztin1} have pointed out that nonlocal effects in a $d$-wave superconductor
can cause an {\em intrinsic} crossover to $T^2$ behaviour in $\lambda(T)$, at $T^\ast
\approx 2 T_{\text c}
\xi_0/\lambda_0$, where $\xi_0$ is the BCS coherence length.  Physically this is because the spatial extent of the Cooper
pair is larger than $\lambda_0$ for a small range of angles around the nodal directions, and affects the field screening at
correspondingly low temperatures.  An estimate using the published value $\xi_0 = 82$~\AA\
\cite{settai}, gives $T^\ast\approx 130$~mK, significantly lower than the value obtained from the fit to our data in
Fig.~\ref{figure2}.  In addition, for a quasi-2D
$d_{x^2 - y^2}$ superconductor such as \cci, nonlocal effects should be very sensitive to geometry, being strongest for
[100] surfaces, vanishing on [110] surfaces, and becoming extremely small for [001] surfaces, where the 2D electronic
structure forces the electrons to move almost parallel to the crystal face.  Our penetration depth probe takes advantage of
this fact, only inducing currents on a [001] surface.  In that case,  $T^\ast$ is determined by the {\em out-of-plane}
coherence length, $\xi_{\text c}  \approx 35$~\AA\ \cite{settai},  with the result that nonlocality should only become a
consideration in our geometry below 56~mK.

In a multi-band metal there is the possibility that superconductivity occurs strongly for only some pieces of the Fermi surface, and is
induced in the others by an internal proximity effect.  This is thought to be relevant to CuO-chain superconductivity in
the cuprates
\cite{xiang96,atkinson96} and to the physics of \sro\ \cite{hiroaki}. In all cases it is expected to lead to a {\em stronger}
$T$-dependence of $\lambda(T)$, due to the presence of  a small temperature scale associated with intrinsically weak
superconductivity in parts of the Fermi surface.  For \cci, which appears to be a $d_{x^2 - y^2}$ superconductor with
line nodes, a proximity effect should result in {\em positive} curvature of $\rho_{\text s}(T)$ \cite{xiang96}, opposite 
to that observed.

The Bose--Einstein condensation theory of Ref.~\onlinecite{chen} predicts
$\Delta\lambda(T) \sim T^{1.5}$ at low $T$, and has been applied to
$\lambda(T)$ measurements on the organic superconductor BEDT \cite{carrington}.  The theory has been developed for
the case of nonoverlapping Cooper pairs in a short-coherence-length, $s$-wave superconductor, as a possible explanation
of the physics of the underdoped cuprates.  It is not likely to be relevant to the case of \cci, which has strongly
overlapping,
$d$-wave pairs.

As none of the scenarios presented so far seems to provide an adequate account of the
$\lambda(T)$ data we propose another possibility, motivated by the NFL properties of the normal state.  There we know
that
$\gamma$, $\chi$ and $m^\ast$ continue to be renormalised down to the lowest temperatures.  If a $T$-dependent
renormalisation also took place {\em within} the superconducting state, what would its effect on $\rho_{\text s}(T)$ be? 
This question is complicated by the fact that the different bands in \cci\ have (field-dependent) cyclotron masses spanning the
range 10 to 100 $m_0$ \cite{settai}, and that
$\rho_{\text s} \propto \langle1/m^\ast\rangle$.  Nevertheless, an effective mass that {\em increased} with decreasing
temperature would introduce the right curvature into $\rho_{\text s}(T)$. To illustrate this point better we have taken our
$\rho_{\text s}(T)$ data, assumed the $d$-wave form for the superfluid density, $n_{\text s}(T) = n_0(1 - \alpha
T/T_{\text c})$, and plotted, below 0.8~K, in the inset of Fig.~3, the temperature dependence of $m^\ast$ for different
values of $\alpha$. In each case
$m^\ast(T)$ shows a cusp-like upturn at low $T$, reminiscent of the quantum-critical behaviour of
$C/T$ in the normal state. Are there other data to support this conjecture? A $T$-dependent renormalisation in the
superconducting state should show up most clearly in the specific heat. In \cci\ the measurement is complicated by the
large low-$T$ entropy of the In nuclei but, when the nuclear contribution is subtracted from the data, $C/T$ still deviates
from the expected $d$-wave $T$-linear behaviour, even showing signs of a low temperature upturn \cite{petrovic,movshovich}.  The
unusual low-$T$ power law in the thermal conductivity, $\kappa(T) \sim T^{3.37}$, might also find a consistent
explanation within this scenario. (As with
$\rho_{\text s}(T)$, $\kappa(T)$ would be dominated by the {\em least} divergent band.)  Further measurements of
thermodynamic properties in the low-$T$ limit should help assess the validity of this proposal.

In conclusion, we have presented the first complete measurements of the London penetration depth of \cci\ in the
low-$T$ limit and observed strong power-law behaviour that confirms the presence of low-lying excitations but is
inconsistent with standard models of $d$-wave superconductivity.  By working with carefully characterised samples and
using a novel resonant-circuit probe that  is inherently insensitive to nonlocal effects and electronic anisotropy we have
been able to rule out extrinsic contributions to the
$\lambda(T)$ signal.  We propose an alternative interpretation, and suggest that the NFL renormalisation occurring in the
normal state of \cci\ might also take place within the superconducting phase, leaving us with the exciting possibility that
\cci\ may be the first example of a quantum-critical superfluid.

We would like to thank D.A.~Bonn, W.N~Hardy, D.E.~Khmelnitskii, P.B.~Littlewood, G.G.~Lonzarich and
N.J.~Wilson for useful discussions.  One of us (DMB) acknowledges the support of Peterhouse, Cambridge.  

More recent measurements from E.~Chia {\em et al.}, which have come to our attention while preparing this work, confirm
$\Delta\lambda(T) \sim T^{1.5}$ at low $T$, using crystals from the same source as ours.


\end{document}